\documentclass[%
 reprint,
superscriptaddress,
nofootinbib,
 amsmath,amssymb,
 aps,
floatfix,
onecolumn,
pre,
longbibliography ]{revtex4-2}

\usepackage{color, soul}
\usepackage{xcolor}
\usepackage{comment}
\usepackage{graphicx}
\usepackage{dcolumn}
\usepackage{bm}
\usepackage{upgreek}

\begin{document}

\preprint{APS/123-QED}

\title{Measurement of Pressure Gradients near the Interface in the Viscous Fingering Instability
}

\author{Savannah D. Gowen}
 \email{gowen22s@uchicago.edu}
 \affiliation{
 Department of Physics and The James Franck and Enrico Fermi Institutes,
 University of  Chicago, Chicago,  IL  60637,  USA}
\author{Thomas E. Videb\ae k} 
    \affiliation{
 Department of Physics and The James Franck and Enrico Fermi Institutes,
 University of  Chicago, Chicago,  IL  60637,  USA}  
    \affiliation{Martin A. Fisher School of Physics, Brandeis University, Waltham, MA 02453, USA}
\author{Sidney R. Nagel} 
    \affiliation{
 Department of Physics and The James Franck and Enrico Fermi Institutes,
 University of  Chicago, Chicago,  IL  60637,  USA}

\date{\today}

\begin{abstract}
The viscous fingering instability, which forms when a less-viscous fluid invades a more-viscous one within a confined geometry, is an iconic system for studying pattern formation. For both miscible and immiscible fluid pairs the growth dynamics change after the initial instability onset and the global structures, typical of late-time growth, are governed by the viscosity ratio. Here we introduce an experimental technique to measure flow throughout the inner and outer fluids. This probes the existence of a new length scale associated with the local pressure gradients around the interface and allows us to compare our results to the predictions of a previously proposed model for late-time finger growth.  
\end{abstract}

\maketitle

\begin{figure}[b!]
\centering
\includegraphics[width=8.6cm]{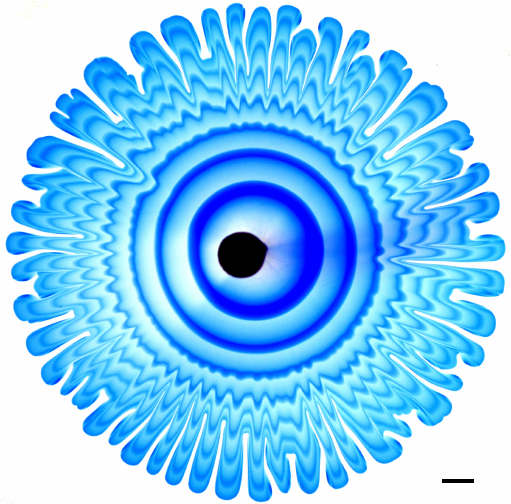}
\caption{
Viscous fingering patterns imaged with alternate injections of dyed and undyed fluid rings. The low-viscosity inner fluid is dyed in alternate rings of blue. The outer fluid is transparent. The scale bar represents 1~cm, $b=420~\upmu\rm{m}$, and $(\eta_{\rm{in}}/\eta_{\rm{out}})=0.067$.
}
\label{Inner_Rings_Color}
\end{figure}

\section{Introduction}
Pattern formation is often seeded by an instability; The selection criteria for the initial pattern shape and length scale have in many cases been well established~\cite{Gollub,Rev_cross,pattern_bk,Rev_langer}. Pattern development at later times has received less attention. Here we consider the patterns formed by the viscous fingering instability – a fluid instability caused by a less viscous fluid (of viscosity $\eta_{\rm{in}}$) displacing a more viscous one (of viscosity $\eta_{\rm{out}}$) within a confined geometry~\cite{homsy,Rev_bensimon,couder,paterson1,paterson2,chuoke}. Pattern formation begins when the initially smooth fluid-fluid interface becomes unstable to the formation of undulations. At onset, the patterns can be characterized by a most unstable wavelength. However, as pattern growth continues experiments have shown that the parameters setting the emergent global structures may be different from those important at onset~\cite{Bischofberger,island}. This demonstrates that the growth dynamics change even after the instability has formed. Here we concentrate on the late-time growth of patterns using a visualization technique, demonstrated in Fig.~\ref{Inner_Rings_Color}, to observe flows throughout the inner and outer fluids.

\begin{figure}[t!]
\centering
\includegraphics[width=6.1in, height=2.1in]{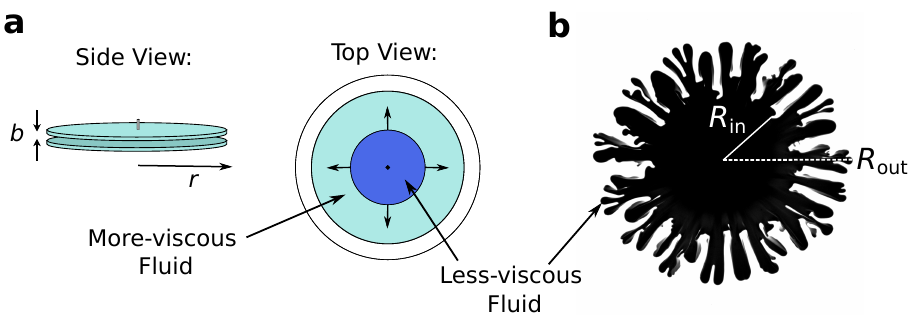}
\caption{
\textbf{(a)} Schematic of a radial Hele-Shaw cell consisting of two circular glass plates separated by a thin uniform gap spacing, $b$. The less-viscous fluid moves radially, displacing the more-viscous one. \textbf{(b)} Finger patterns of late-time growth viewed from above. The inner fluid is dyed and the outer fluid is left transparent for contrast. Characteristic length scales $R_{\rm{in}}$ and $R_{\rm{out}}$ are labeled.
}
\label{Hele-Shaw}
\end{figure}

Viscous fingering can be observed in pairs of fluids, with viscosity difference $\Delta \eta \equiv \eta_{\rm{out}} - \eta_{\rm{in}}$, confined in a Hele-Shaw cell consisting of two parallel glass plates separated by a uniform gap of width $b$, as illustrated in Fig.~\ref{Hele-Shaw}a. For pairs of immiscible fluids, Saffman and Taylor ~\cite{SaffT} showed that perturbations at the interface between a less-viscous invading fluid and a more-viscous one will grow with a critical wavelength:  $\lambda_{c}= \pi b (\sigma/\Delta \eta V)^{1/2}$ where $\sigma$ is the interfacial tension and $V$ is the velocity of the interface. In the case of miscible fluids, where $\sigma$ is negligible, the most unstable wavelength varies as $ \lambda_{c} \propto C b $ ~\cite{paterson2,lajeunesse,nagel_gal,kim_homsy,dias}, where $C$ is a constant that has been shown to depend weakly on the viscosity ratio $(\eta_{\rm{in}}/\eta_{\rm{out}})$~\cite{Thomas}. Despite these differences, for both miscible and immiscible pairs of fluids, viscous fingering produces similar global structures: the patterns have an inner stable radius, $R_{\rm{in}}$, characterized by the inner circular region where the outer fluid has been completely displaced and an outer radius, $R_{\rm{out}}$, which contains the entire interface as shown in Fig.~\ref{Hele-Shaw}b. 

A detailed investigation of this late-time growth~\cite{Bischofberger} showed that for patterns with the same $\lambda_{c}$, the inner stable radius grows faster with increasing $(\eta_{\rm{in}}/\eta_{\rm{out}})$, independent of surface tension and qualitatively independent of the geometry of the interface in the third dimension between the plates.  In an attempt to account for the emergence of the viscosity ratio as a control parameter for these ubiquitous global features, that paper also proposed a model to describe the late-time growth of the fingers once they had formed.  The key assumptions of that model are two length scales related to the decay of local pressure gradients from the interface into the inner and outer fluids. The experimental technique introduced here allows direct access to the pressure gradients around the interface, and can therefore be used to test these assumptions.

We conduct experiments in a radial Hele-Shaw cell as shown in Fig.~\ref{Hele-Shaw}a and described in detail in the Methods section. Miscible fluids are injected at a constant flow rate through a small hole in the top plate. The outer fluid is injected first followed by the less viscous invading fluid. To visualize flow throughout the cell we create dyed and clear versions of both the inner and outer fluids, being careful that the addition of dye to either fluid preserves both its density and viscosity. We then alternately inject small volumes of, \textit{e.g.}, the inner fluid in contrasting colors such that dyed volumes of fluid can be tracked as they propagate outward radially as rings. Whereas the interface is commonly tracked by using inner and outer fluids with contrasting colors as was shown in Fig.~\ref{Hele-Shaw}b, alternate injection of dyed and clear fluid allows contrast for observing flow anywhere in the cell and does so without perturbing the flow. 

In a confined flow, the fluid velocity  $\langle \vec{u}\rangle$ is proportional to the pressure gradient, $\vec{\nabla} P$, as given by Darcy’s Law: 

\begin{equation}
 \langle \vec{u}\rangle =  -\frac{\kappa}{\eta} \vec{\nabla} P 
 \label{darcy}
\end{equation}
where $\eta$ is the fluid viscosity and $\kappa$ is the permeability of the medium.  Thus by tracking dyed fluid rings we can measure the local flow velocity and obtain the local pressure gradients. The technique, the details of which are described in the \textit{Methods} section, can be used separately in either the inner [Fig.~\ref{Inner_Rings_Color}] or outer fluid or in both fluids simultaneously [Fig.~\ref{label_scheme}a].

Our experiments demonstrate a non-invasive method for measuring the pressure gradients as a function of angle in the neighborhood of the interface between the fluids. We show that these pressure gradients decay with a characteristic distance behind and beyond the undulating interface and thus determine two new length scales associated with the fingering instability. In the \textit{Results} section, we use this technique to measure the local pressure gradients in both the inner and outer fluids as a function of distance relative to the moving interface; we extract the decay lengths and analyze them in terms of the experimental variables.  In the \textit{Discussion} section, we relate these lengths to those that appeared as hypothesized free parameters in the model for late-time interface evolution~\cite{Bischofberger}. We compare our results with the model predictions and propose modifications that agree well with experimental measurements.

\section{Results}

\subsection{Detection of Local Pressure Gradients}

\begin{figure*}[tb!]
\centering
\includegraphics[width=6.8in,height=2.8in]{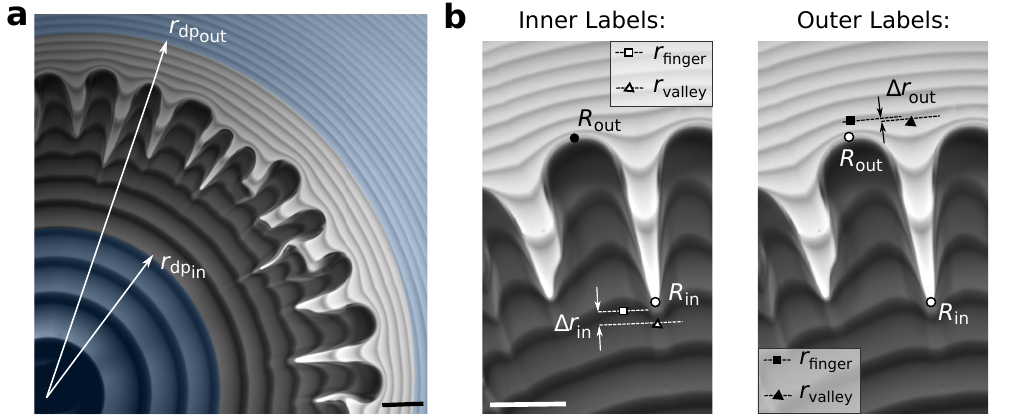}
\caption{Length scales of viscous fingering patterns. \textbf{(a)} One pattern quadrant is shown that has been imaged using alternate dyed-ring injection in both the inner and outer fluids.  A region surrounding the interface over which local pressure gradients perturb flow is estimated with the bounds  $r_{\rm{dp_{\rm{in}}}}$ and $r_{\rm{dp_{\rm{out}}}}$. \textbf{(b)} Dyed rings are analyzed in local pattern segments pertaining to a finger and an adjacent valley. Length scales $r_{\rm{finger}}$ and $r_{\rm{valley}}$ along a single dyed ring are labeled in the inner and outer fluids. The scale bars represent 0.5~cm.
}
\label{label_scheme}
\end{figure*}

Using our experimental technique, we observe that as rings of dyed fluid approach the unstable interface there is a finite distance at which undulations form along the rings. Flow in this system is driven by an imposed pressure gradient between the inlet pressure and the atmospheric pressure of the air surrounding the open cell. In a stable system, alternately injected rings of dyed fluid would remain circular. The variations in the local velocity field that create the observed undulations reveal the presence of local pressure gradients and a length scale over which they influence flow. In Fig.~\ref{label_scheme}a, we estimate the region surrounding the interface over which local pressure gradients perturb the flow, bounded in the inner fluid by $r_{\rm{dp_{\rm{in}}}}$, and in the outer fluid by $r_{\rm{dp_{\rm{out}}}}$.

\subsection{Local Pressure Gradient Decay in the Inner Fluid}

To quantify the length scale associated with the local pressure gradient behind the interface, we measure the magnitude of undulations along a single dyed fluid ring. As shown in Fig.~\ref{label_scheme}b, we identify the radial location of the dyed ring behind a finger protrusion as $r_{\rm{finger}}$, and behind an adjacent valley as $r_{\rm{valley}}$; we calculate the amplitude of an undulation, $\Delta r_{\rm{in}}\equiv r_{\rm{finger}}-r_{\rm{valley}}$. These measurements are taken instantaneously and are repeated as parameters evolve with the introduction of more inner fluid volume into the cell. Measurements along a ring terminate when the fluid arrives at the interface. We repeat this procedure behind several fingers.

\begin{figure}[htbp!]
\centering
\includegraphics[width=3.4in,height=2.8in]{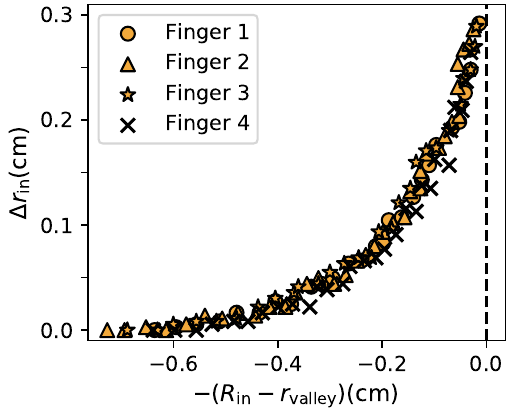}
\caption{
Undulation magnitude versus the negative distance behind the interface for a single dyed ring. The interface, $R_{\rm{in}}$, is marked by the dashed vertical line. Deformations are measured behind four different fingers. Experimental parameters are: $b=420~\mathrm{\upmu m}$ and $(\eta_{\rm{in}}/\eta_{\rm{out}})$ = 0.104. 
}
\label{onering}
\end{figure}

\begin{figure*}[htbp!]
\centering
\includegraphics[width=6.8in,height=3.5in]{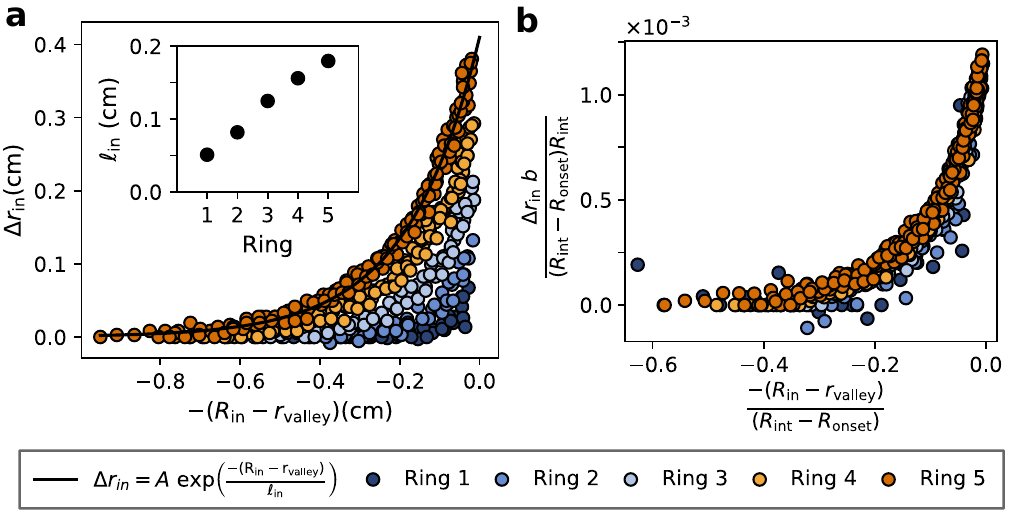}
\caption{
Characterization and scaling of perturbation growth in the inner fluid.  \textbf{(a)} The magnitude of undulations versus distance behind the interface for five dyed rings labeled in order of their relative injection times. Curves are fit by an exponential function to determine a characteristic decay length, $\ell_{\rm{in}}$. The inset shows $\ell_{\rm{in}}$ increasing as a function of ring injection sequence (time). \textbf{(b)} The x- and y-axes are re-scaled such that the deformation curves from sequential rings collapse onto a single curve. Experimental parameters are: $b=420~\mathrm{\upmu m}$ and $(\eta_{\rm{in}}/\eta_{\rm{out}})$ = 0.104. 
}
\label{allrings}
\end{figure*}

Despite local differences in the sizes and shapes of the interfacial fingers, the resulting deformations develop similarly for fingers that onset at approximately the same time.  Figure~\ref{onering} shows the magnitude of undulations, $\Delta r_{\rm{in}}$, behind four fingers along a single dyed ring versus the negative distance behind the interface $-(R_{\rm{in}}-r_{\rm{valley}})$.  The zero on the x-axis corresponds to the location of the stable interface radius, $R_{\rm{in}}$. Dyed rings far from the interface begin with zero deformations (long-wavelength perturbations appearing farther behind the interface can lead to artificially negative deformation values, which we subtract away). $\Delta r_{\rm{in}}$ grows with the fluid's proximity to the interface in a similar way for different fingers.

We repeat these measurements for the same four fingers but along rings injected at different times throughout the pattern formation. Figure~\ref{allrings}a shows these deformation curves gathered from five dyed fluid rings. Fitting these curves to an exponential form $\Delta r_{\rm{in}}=A \exp [-(R_{\rm{in}}-r_{\rm{valley}})/\ell_{\rm{in}}$] shows good agreement with the data. The associated length scale, $\ell_{\rm{in}}$, is shown as a function of ring injection sequence in the inset. $\ell_{\rm{in}}$ increases with time, showing the growth of local pressure gradients from the interface into the inner fluid.

To collapse data collected from different dyed rings, we examined many combinations of system parameters to re-scale the x- and y-axes. These included the radial locations associated with the tips and valleys of the dyed rings in the bulk; $r_{\rm{finger}},\ r_{\rm{valley}}$; the radial locations of the interface (measured locally) and its average radius: $R_{\rm{in}}$, $R_{\rm{out}}$, $R_{\rm{int}}\equiv \frac{(R_{\rm{in}}+R_{\rm{out}})}{2}$; the radius at which the interface initially became unstable for a finger, $R_{\rm{onset}}$; as well as the length, width, and tip diameter of individual fingers. The re-scaled axes must, in the end, be dimensionless.  Our best re-scaling of these curves is shown in Fig.~\ref{allrings}b where the x-axis is scaled by $R_{\rm{int}}-R_{\rm{onset}}$ and the y-axis by $(R_{\rm{int}}-R_{\rm{onset}})R_{\rm{int}}/b$. This scaling yields a single master curve in which each data point varies in time and space. We note that $R_{\rm{int}}$ may have also been substituted by $R_{\rm{in}}$ with little change to the result.  The collapse of this data suggests a form of the characteristic decay length given by the denominator of the re-scaled x-axis: $ \ell_{\rm{in}} \propto (R_{\rm{int}}-R_{\rm{onset}})$.

\begin{figure*}
\centering
\includegraphics[width=6.7in,height=3.1in]{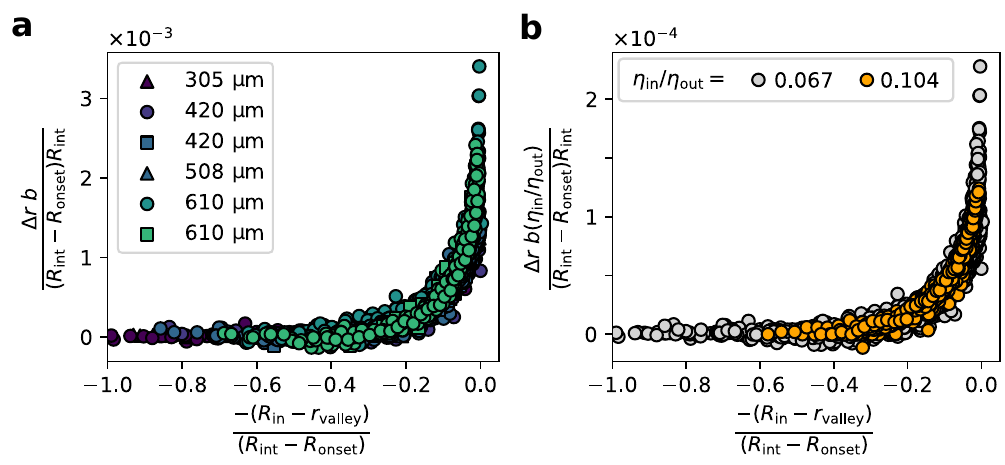}
\caption{
Scaled collapse of inner fluid experiments with varied control parameters. \textbf{(a)} Scaled curves from six experiments for $(\eta_{\rm{in}}/\eta_{\rm{out}})=0.06$ and varied gap widths, $305~\mathrm{\upmu m} \le b \le 610~\mathrm{\upmu m}$. \textbf{(b)} The y-axis is scaled by $(\eta_{\rm{in}}/\eta_{\rm{out}})$ to collapse data from experiments with different viscosity ratios. Here, $b=420~\upmu\rm{m}$.
}
\label{All Experiments}
\end{figure*}

We repeat this scaling procedure for data collected from experiments with varied gap widths to produce the collapsed curve shown in Fig.~\ref{All Experiments}a (in this analysis we have limited ourselves to a regime in which the finger length is significantly larger than the finger width). These results are consistent with the decay length of local pressure gradients, $\ell_{\rm{in}}$, being independent of the gap spacing $b$. 

We also examine the relation between the pressure-gradient decay and the viscosity ratio, $(\eta_{\rm{in}}/\eta_{\rm{out}})$.  Figure ~\ref{All Experiments}b shows that data from a higher viscosity ratio experiment collapses onto that shown in Fig.~\ref{All Experiments}a when the y-axis is further scaled by $(\eta_{\rm{in}}/\eta_{\rm{out}})$. This re-scaling does not affect the x-axis; we conclude that the decay length is independent of the viscosity ratio, although the maximum deformation size is sensitive to it.

We fit the scaled data from all these experiments to extract a form for the characteristic decay length of local pressure gradients in the inner fluid, given by: 
\begin{equation}
\label{eqn:l_in}
\ell_{\rm{in}} \approx (0.09 \pm 0.01)(R_{\rm{int}}-R_{\rm{onset}}).
\end{equation}
We interpret this length as a penetration depth for local pressure gradients extending behind the interface. It depends on the radial distance traveled by the interface since the initial onset of the instability.

\subsection{Local Pressure Gradient Decay in the Outer Fluid}

We perform an analysis for the outer fluid by measuring the undulations on dyed fluid rings \textit{in front} of the interface. Similar to our treatment of the inner fluid, the magnitude of an undulation, $\Delta r_{\rm{out}}$, is determined by taking the difference between the radius of a dyed ring in front of a finger, $r_{\rm{finger}}$, and in front of the adjacent valley, $r_{\rm{valley}}$, as demonstrated in Fig.~\ref{label_scheme}b. Undulations decay with the distance from the interface, $ -(R_{\rm{out}}-r_{\rm{finger}})$, now defined with respect to the outer radius, $R_{\rm{out}}$. It is possible for the curves to plateau to positive or negative values rather than zero due to long-wavelength effects. To account for this, we apply a small vertical shift so that all curves plateau to zero far from the interface.

\begin{figure}[tb!]
\centering
\includegraphics[width=6.75in,height=3in]{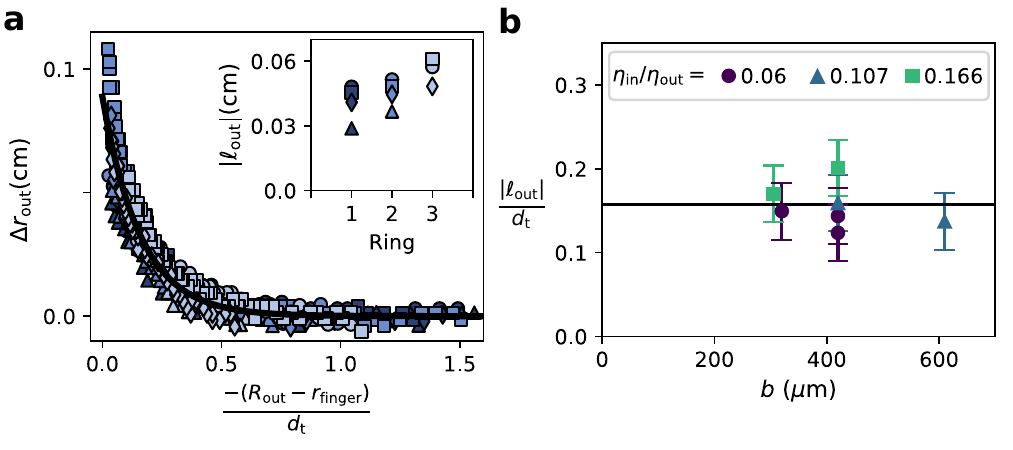}
\caption{
Perturbation decay in the outer fluid. \textbf{(a)} Deformation curves from three sequential dyed rings (color-coded as in inset) ahead of four different fingers (represented by square, diamond, circle, and triangle) collapse when the x-axis is scaled by the local finger tip diameter, $d_{\rm{t}}$. The scaled master curve is fit to an exponential form. The inset shows individual fits for $| \ell_{\rm{out}} |$ as a function of ring injection sequence before scaling. Experimental parameters are: $b=420~\mathrm{\upmu m}$ and $(\eta_{\rm{in}}/\eta_{\rm{out}})$ = 0.107. \textbf{(b)} The normalized decay length $ | \ell_{\rm{out}} |/ d_{\rm{t}}$ from different experiments is plotted versus gap width, $b$. Different color-symbol combinations represent different viscosity ratios. Individual experimental values agree with the overall fit (represented by the horizontal line).
}
\label{of_exp_collapse}
\end{figure}

Similar to the inner fluid, deformation curves are well characterized by an exponential function of the form, $\Delta r_{\rm{out}}=A \exp [(-(R_{\rm{out}}-r_{\rm{finger}})/ \ell_{\rm{out}}]$. Curves associated with undulations along three dyed rings and ahead of four different fingers are fit to extract the characteristic decay length, $| \ell_{\rm{out}} |$, shown in the inset of Fig.~\ref{of_exp_collapse}a. $| \ell_{\rm{out}} |$ increases in time for each sequential ring but, compared to the inner fluid, shows greater variability between fingers along a single ring. This variability is reduced by re-scaling the x-axis of the deformation curves. Our best collapse of the data is shown in Fig.~\ref{of_exp_collapse}a where we scale by the local diameter of a protruding finger, $d_{\rm{t}}$, as characterized by fitting a circle to the tip of the finger. All undulations appear to converge to a similar maximum amplitude without scaling the y-axis. Collapsed data from the experiment is fit to an exponential form to extract the normalized decay length $ \ell_{\rm{out}} /d_{\rm{t}}$.

 We repeat this analysis by collecting deformation curves along multiple rings and behind several fingers for experiments with varied gap widths and viscosity ratios.  Data from all experiments collapse best when we re-scale the y-axis by dividing by the gap width, $b$, and multiplying by the viscosity ratio, $(\eta_{\rm{in}}/\eta_{\rm{out}})$. However, we note that other scaling parameters could not be ruled out within the noise of this data; for example, the y-axis could have been re-scaled by the finger tip diameter, $d_{\rm{t}}$, but this introduced slightly more noise to the data collapse. Also, scaling the x-axis by $R_{\rm{int}}-R_{\rm{onset}}$ yielded a slightly worse collapse, but one still plausible within the noise. 

Figure~\ref{of_exp_collapse}b shows the magnitude of the normalized decay length, $ | \ell_{\rm{out}} |/ d_{\rm{t}}$, versus the gap width, $b$, and the viscosity ratio, $(\eta_{\rm{in}}/\eta_{\rm{out}})$, for individual experiments. The fit from the collapse of all the data is marked by the horizontal line. The decay length $ | \ell_{\rm{out}} |/ d_{\rm{t}}$ appears constant, and thus independent of control parameters $b$ and $(\eta_{\rm{in}}/\eta_{\rm{out}})$. From the fit and x-axis scaling relation, we can extract a functional form for the outer fluid pressure-gradient penetration depth:
\begin{equation}
\label{eqn:l_out}
\lvert \ell_{\rm{out}} \rvert \approx (0.16 \pm 0.02) d_{\rm{t}}.
\end{equation}

These experiments and analyses have allowed us to determine and characterize a new pressure-gradient-decay length scale associated with the late time growth of patterns. It should be noted that the dynamics in the inner and outer fluids are different from each other; the same non-dimensional x- and y-axes used in the inner fluid do a poor job of collapsing the data in the outer fluid. Likewise, the outer-fluid scaling relation gives a poor collapse for data in the inner fluid. This might be explained by differences in flow structure within the gap for the inner and outer fluids, as we discuss in the following section.

\section{Discussion}

Our experiments show that local pressure gradients decay exponentially from the interface. A number of earlier studies have examined the non-linear growth of viscous fingering patterns~\cite{swinney,Tan,chen,booth,lehman,zimmerman,juanes,Zim_homs,Bischofberger,island,cheng_nagel,zheng_stone,vaquero,oliveira_meiburg,Rev_bensimon}, but only a few have endeavored to analyze pressure and local velocity gradients more explicitly ~\cite{miranda,holzbecher,weitz}. We were unable to find a direct comparison between the observations of this literature and the pressure-gradient-decay length that we have reported here. Nevertheless, a comparison can be made between our findings and the analytical form of pressure perturbations proposed in the linear stability analysis. Exponential perturbation decay is often assumed in the rectilinear geometry~\cite{couder,SaffT}; however, in the radial geometry, a power-law form of $r$ is more common~\cite{paterson2}. In either case, the decay of perturbations is related to the most-unstable wavelength. We see this in our outer fluid where the decay length is proportional to the diameter of a finger, which is related to the wavelength. Notably, the decay in the inner fluid does not show the same dependence. 

\subsection{Comparison to model}
A direct comparison can be made between our data and the model for the local pressure drop between two adjacent uncoupled fluid channels proposed by Bischofberger \textit{et. al.}~\cite{Bischofberger}. Figure~\ref{cartoon} illustrates two neighboring radial flow channels -- one with a finger protrusion whose interface radius is marked as $R_{\rm{out}}$, and the other an adjacent valley with the interface marked as $R_{\rm{in}}$. While the net pressure drop across both channels is identical, a local pressure difference near the interface leads to a faster interface velocity for the finger, $V_{\rm{out}}$, than for the valley, $V_{\rm{in}}$.  The model distinguishes the range over which these local pressure gradients influence velocity by defining radial positions in the inner and outer fluids, $r_{\rm{dp_{\rm{in}}}}$ and $r_{\rm{dp_{\rm{out}}}}$ respectively, as bounds where the local pressures become equal in both channels. Flow is assumed to be entirely radial and to obey Darcy's law [Eq.~\ref{darcy}] such that the interface velocity is governed by the pressure drop in each channel. This yields an expression for the ratio of the outer to inner interface velocity:

\begin{figure}[b!]
\centering
\includegraphics[width=3.4in,height=3in]{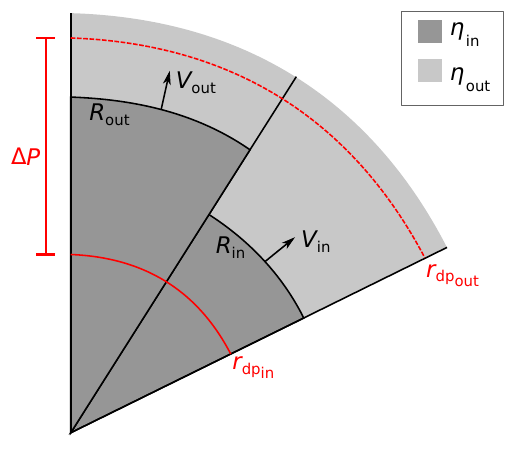}
\caption{
Two independent radial flow channels corresponding to a finger and adjacent valley. The interface is denoted by $R_{\rm{out}}$ in the finger and by $R_{\rm{in}}$ in the adjacent valley. The net pressure drop across each channel is the same, however, the \textit{local} pressure gradients are different around the interface: flow is identical in both channels up to $r_{\rm{dp_{\rm{in}}}}$ and beyond $r_{\rm{dp}_{out}}$. 
}
\label{cartoon}
\end{figure}

\begin{equation}
\label{eqn:model}
\frac{V_{\rm{out}}}{V_{\rm{in}}} = \frac{R_{\rm{in}}}{R_{\rm{out}}}\bigg[\frac{(\eta_{\rm{in}}/\eta_{\rm{out}})\ln(\frac{R_{\rm{in}}}{r_{\rm{dp_{\rm{in}}}}})+\ln(\frac{r_{\rm{dp_{\rm{out}}}}}{R_{\rm{in}}})}{(\eta_{\rm{in}}/\eta_{\rm{out}})\ln(\frac{R_{\rm{out}}}{r_{\rm{dp_{\rm{in}}}}})+\ln(\frac{r_{\rm{dp_{\rm{out}}}}}{R_{\rm{out}}})} \bigg].
\end{equation}
This model predicts the flow after the initial instability onset and has the viscosity ratio, $(\eta_{\rm{in}}/\eta_{\rm{out}})$, appearing as a control parameter. The parameters $r_{\rm{dp_{\rm{in}}}}$ and $r_{\rm{dp_{\rm{out}}}}$ were introduced as free parameters. Our experiments validate their existence and provide a functional characterization for each. 

\begin{figure*}

\centering
\includegraphics[width=6.75in,height=2.9in]{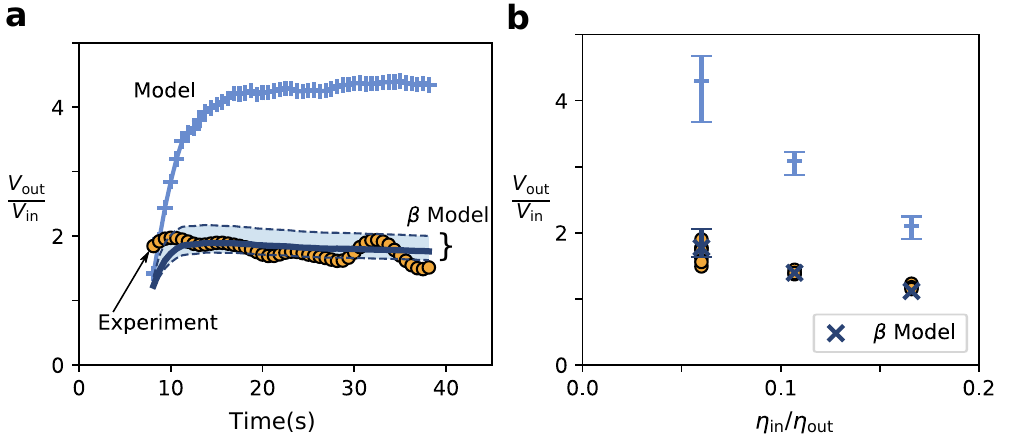}
\caption{
Comparison of the measured and predicted velocity ratio. (\textbf{a}) The velocity ratio, ($V_{\rm{in}}/V_{\rm{out}}$), for a single finger-valley pair is shown over the course of an experiment and is compared to model predictions. Model data (connected plus symbols) is shown to differ significantly from experimental data (circles) unless modified to account for flow structure in the gap (line with the shaded region). Values for the $\beta$ model assume that the equal-pressure point behind and ahead of the interface occurs where ring undulations decay to 90\% of their maximum value; the bounds of the shaded region correspond to 75\% (upper) and 95\% (lower) thresholds. Data is shown for: $b=420~\upmu \rm{m}$, $(\eta_{\rm{in}}/\eta_{\rm{out}})$ = 0.067. (\textbf{b}) The average plateau value for ($V_{\rm{out}}/V_{\rm{in}}$) (such as that seen in (a)) is plotted versus $(\eta_{\rm{in}}/\eta_{\rm{out}})$. The experiment and model outputs are marked with the same symbols as in (a). For the model, we plot an average of outputs for 3-4 fingers with the bar denoting the range from smallest to largest values. The averaged $\beta$ adjusted model output is marked by a cross. Uncertainty due to finger variation and parameter definitions are smaller than the symbol size for the two larger viscosity ratios and are thus omitted. The error bars shown correspond to the bounds of the shaded region in (a). All experiments were done with $b=420~\mathrm{\upmu m}$.
}
\label{model_compare}
\end{figure*}

We compare the model's predictions with our experiments for a single finger protrusion by analyzing our data in a manner analogous to the model. We measure the interface velocities of a finger tip and its adjacent valley; these are the gradients of $R_{\rm{out}}$ and $R_{\rm{in}}$ as a function of time. Figure~\ref{model_compare}a shows the ratio of the velocities, $(V_{\rm{out}} / V_{\rm{in}})$, versus time once the finger has reached a steady growth rate, thus excluding a region at the beginning of pattern growth where the velocity ratio rises rapidly.

To compare our experiments to Eq.~\ref{eqn:model}, we determine $r_{\rm{dp_{\rm{out}}}}$ and $r_{\rm{dp_{\rm{in}}}}$ with respect to $\ell_{\rm{out}}$ and $\ell_{\rm{in}}$ ($i.e.$, the parameters at which local pressure gradients from the interface have ceased to affect the flow) determined experimentally and given by Eqs.~\ref{eqn:l_in} and~\ref{eqn:l_out}. Here we choose the length at which undulations have decayed to 90\% of their value at the interface, $2.3 \ell$, to characterize the model parameters:

\begin{equation}
r_{\rm{dp_{\rm{in}}}} \equiv R_{\rm{in}}-2.3\ell_{\rm{in}}
\end{equation}

 and

\begin{equation}
r_{\rm{dp_{\rm{out}}}} \equiv R_{\rm{out}}+2.3\lvert\ell_{\rm{out}}\rvert.
\end{equation}
Using these, and the other parameters determined experimentally, in Eq.~\ref{eqn:model} yields a predicted value for the velocity ratio shown by the connected plus signs in Fig.~\ref{model_compare}a. The velocity ratio predicted by the model is almost double the experimental value (circles).

To address the discrepancy between the model and experiment we apply a correction that takes into account the fact that there are three stratified layers of miscible fluid in the gap: a thin layer of outer fluid adjacent to both the top and bottom plates, and a tongue of inner fluid in the center~\cite{videbaek2020,lajeunesse}. Ahead of the interface, there is only a single layer; behind the interface, the inner fluid tongue occupies only a fraction, $\beta$, of the gap.  We average across the gap, incorporating the stratified nature of the fluids, to find an effective viscosity ratio:

 \begin{equation}
\left(\frac{\eta_{\rm{in}}}{\eta_{\rm{out}}}\right)^{*} = \frac{\eta_{\rm{in}}}{\eta_{\rm{out}}}\bigg(\frac{1}{(\eta_{\rm{in}}/\eta_{\rm{out}})+\beta^{3}(1-(\eta_{\rm{in}}/\eta_{\rm{out}}))}\bigg).
\end{equation}
We measure $\beta$ experimentally as a function of viscosity ratio and use this averaged viscosity ratio, $(\eta_{\rm{in}}/\eta_{\rm{out}})^{*}$, in Eq.~\ref{eqn:model}. This yields the curve shown by the dark blue line in Fig.~\ref{model_compare}a, which we will refer to as the $\beta$-adjusted model. The $\beta$-adjusted model corresponds well with the average magnitude and trend of the data. 

We note that the greatest source of uncertainty in the model prediction comes from the factor used in the definition of model parameters $r_{\rm{dp_{\rm{out}}}}$ and $r_{\rm{dp_{\rm{in}}}}$. The model defines a \textit{finite} length behind and in front of the interface over which local pressure gradients decay, while in the experiment we observe a smooth exponential decay. We chose to compare the model length scales with respect to the experiment as the length at which undulations have decayed to 90\% of their maximum value. In Fig.~\ref{model_compare}a the shaded region bounded by dashed lines represents the range of model predictions had we chosen this factor to correspond to a decay range between 75\% and 95\%. The inclusion of this error does not change the conclusions of our results. Including other sources of error (such as the uncertainty in the measured value of $\beta$, $\ell_{in}$, or $\ell_{out}$) does not significantly increase this spread.

This comparison is repeated for multiple fingers and for experiments at varied viscosity ratios. At late times the velocity ratio plateaus to a nearly constant value, as demonstrated in Fig.~\ref{model_compare}a. In Fig.~\ref{model_compare}b, we show the average value of the plateau region in experiments for different control parameters and find that the experimental velocity ratio (circles) decreases with increasing viscosity ratio, $(\eta_{\rm{in}}/\eta_{\rm{out}})$. This trend is also captured by the model.

The correction for the stratification of the flow is necessary to produce a good quantitative fit to the data for all of the experiments. This quantitative comparison also works for larger gap widths. However, at smaller gap widths, the comparison deteriorates as azimuthal flows become more prominent in the pattern formation, as evidenced by tip-splitting events and channel merging. These effects will also emerge for the gap widths shown, but at larger radii than observed here.

\section{Conclusions}

The experiments reported here have shown that the fluid velocities, and therefore the pressure gradients, throughout a Hele-Shaw cell can be determined by a technique in which the fluid is differentiated by alternate injection of dyed fluid.  Using this technique, we have identified a radial decay length for the local pressure gradient, extending both in front of and behind the propagating interface.  Our characterizations of the local pressure-gradient decay lengths in the inner and outer fluid, $\ell_{\rm{in}}$ and $\ell_{\rm{out}}$, represent the best fits to our data. As indicated in the \textit{Results} section, other scaling choices, within the noise and fluctuations of the data, could have been made but would not have significantly changed our conclusions.

The existence of a length scale related to the decay of local pressure gradients was proposed to relate late-time pattern formation to the control parameter, $(\eta_{\rm{in}}/\eta_{\rm{out}})$~\cite{Bischofberger}. Despite the simplifications made in constructing that model, our experiments show that it captures the local dynamics of pattern formation and the dependence of the velocity ratio on the viscosity ratio as a control parameter surprisingly well. It is predictive of the magnitude of the velocity ratio when a correction is made for the stratified flows behind the interface. The need for such a correction shows that fluid-structure within the third dimension of the gap is important to pattern development. 

While the results presented here were for viscous fingering with miscible fluids, the technique of using alternately dyed rings should be adaptable to immiscible fluids (as well other systems such as colloidal or non-Newtonian fluids~\cite{colloids,non_newt}). The stratified flow behind the interface in miscible fluids led to important differences between how the inner and outer fluids were treated.  For example, we saw that the predictions of the model for the velocity ratio did not correspond quantitatively with the data unless the effect of stratification was included. Therefore, it is perhaps not surprising that the decay lengths of the inner and outer fluids scale differently. In the case of immiscible fluids, where only a very thin layer of the outer fluid remains at the plate surfaces~{\cite{libchaber}}, the stratification of the flow is much less significant. In that case the correspondence between the model and experiment might be sufficient without any corrections.

\section*{Methods}

\subsection{Apparatus}

Experiments are performed in a radial Hele-Shaw cell apparatus consisting of two flat, circular glass plates of either 1.9~cm or 1.7~cm thickness and a diameter of 28~cm. The two plates are separated by a uniform gap thickness using spacers which range in size from $ 305~\mathrm{\upmu m} - 610~\mathrm{\upmu m}$. Fluid is injected through a 1.6~mm hole in the center of the top plate and propagates outward radially at a fixed volumetric flow rate \textit{Q}=2~ml/min. Pattern formation is recorded at a typical frame rate of 8 f.p.s. using a Prosilica GX3300 camera. A blue light filter is used to enhance contrast.

The fluids used in these experiments are miscible mixtures of glycerol and water. Tuning the concentration of glycerol to water allows us to vary the fluid viscosity between 1-1100~cP. We measure fluid viscosity using an SVM 3001 viscometer. Each fluid exists in an optically transparent version and a dyed version. Dyed fluid is colored using Brilliant Blue G dye from Alfa Aesar in a concentration of 0.08~mg/100~ml. 

\subsection{Alternate Fluid Injection Technique}

In these experiments, we exploit the nature of the Hele-Shaw flow geometry to isolate fluid velocity at the cell’s mid-plane without adding any tracer other than molecular scale dye. Our technique allows us to track flow velocity throughout the cell and thus indirectly determine local pressure gradients by observing variations in the velocity field. The confinement of the Hele-Shaw cell renders traditional flow-tracking techniques like particle image velocimetry (PIV) ineffective. 

We alternately inject a dyed and clear version of the same fluid continuously into the cell using two syringe pumps (NE-1000 from New Era Pump Systems Inc.). Each syringe pump is connected via a TTL input to an SFG-205 function generator (Global Specialties) which powers the pumps on and off using a square wave signal. A T-junction joins separate feed tubing for dyed and undyed fluid at the inlet of the cell. We set a duty cycle of 70\% so that there is slightly more undyed than dyed fluid and vary the frequency of injection from 100~mHz to 500~mHz in order to better resolve the dyed fluid volumes. Small volumes of dyed fluid will propagate outward from the inlet as rings of fixed volume but with an increasing circumference. Injecting many of such dyed volumes allows us to track dyed fluid both spatially and temporally.

\subsection{Alternate Injection Concept: Measuring Flow Velocity at the Mid-plane of the Cell's Gap}

 The nature of Poiseuille flow in the cell's gap allows us to relate the velocity of the dyed fluid ring as measured from above the cell with the velocity along the mid-plane of the cell's gap.

 \begin{figure}[t!]
\includegraphics[width=6in,height=1.6in]{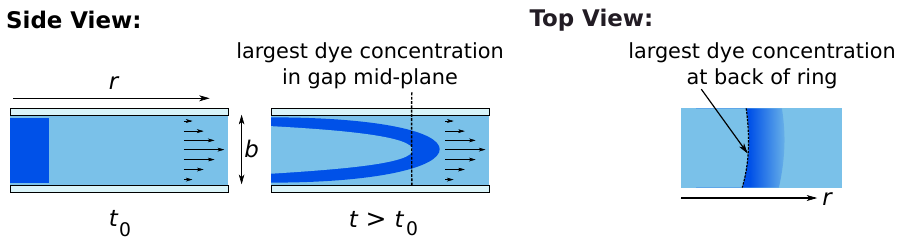}
\caption{
 An illustration of a small volume of dyed fluid viewed between two glass plates stretched by the steady-state Poiseuille flow profile. The black arrows in the gap denote the velocity as a function of height. Viewed from the side, the $r$ location of the highest gap averaged concentration corresponds to the boundary between the back of the dyed ring and the tip of the incoming undyed fluid, which lies in the middle of the gap. Here the aspect ratio is highly exaggerated, as $b \ll r$. Viewed from above, the highest concentration of dye appears around the back of a dyed ring.  
}
\label{ringcartoon}
\end{figure}

 \begin{figure}[b!]
\includegraphics[width=5.5in,height=1.6in]{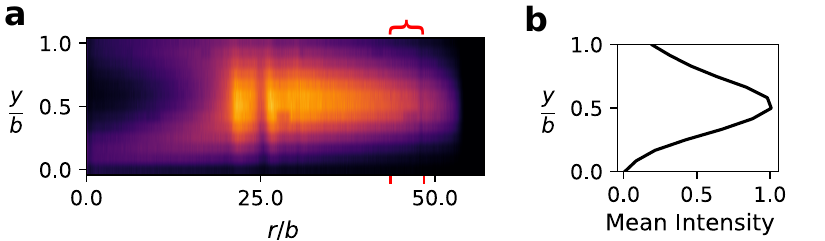}
\caption{Confocal imaging of Poiseuille flow profile.  (\textbf{a}) Fluorescently dyed fluid is injected alternately with undyed fluid between two glass plates separated by gap width $b \approx 533~\mathrm{\upmu m}$. The fluid is imaged with a confocal microscope which samples slices at 12 heights, $y/b$, within the gap. Slices are vertically stacked to reconstruct the fluid profile. The aspect ratio has been exaggerated as $y \ll r$. (\textbf{b}) The mean intensity is averaged over gap height, $y/b$, for a radial segment in the image in (a) as marked by the red bracket and tick marks. The highest mean pixel intensity is located at the center of the gap where dyed fluid is most concentrated, as expected of Poiseuille flow.  The fluid used has a viscosity of $\eta = 31.0~\mathrm{cP}$. 
}
\label{confocal}
\end{figure}

 Figure~\ref{ringcartoon} shows a schematic illustration of the morphology of a small volume of dyed fluid surrounded by clear fluid propagating through the gap between two glass plates. The fluid is stretched into a parabola by the Poiseuille flow due to no-slip boundary conditions with the plates. As viewed from above the plates, the highest concentration of dye appears at the back of the dyed fluid ring. As viewed within the gap, this region of the highest concentrated dye corresponds to where the dyed fluid is displaced by the tip of the trailing undyed fluid, which is located at the mid-plane of the cell's gap. Thus by tracking where dye appears most intensely concentrated at the back of the fluid ring, we can isolate flow velocity at the mid-plane of the cell's gap. We have verified qualitatively the assumption of Poiseuille flow within the gap using a Caliber I.D. RS-G4 confocal microscope to image a fluorescent-dyed ring of fluid injected between two glass plates, as seen in Fig.~\ref{confocal}. Figure~\ref{confocal}b shows that the highest dye intensity, and thus largest concentration of dye, is found in the mid-plane of the gap.   We also compare the squared radial trajectory of a single dyed fluid ring injected into undyed fluid of the same viscosity in our Hele-Shaw cell with that theoretically predicted for Poiseuille flow in the gap's mid-plane. We find an agreement to within 5\%. We assume that the highest dye concentration corresponding to fluid flow at the mid-plane of the gap is also the case for stratified flow.

\section*{Acknowledgements}

We would like to thank Samar Alqatari, Severine Atis, Irmgard Bischofberger, Rudro Biswas, Justin Burton, and Zhaoning Liu for insightful discussions. This work was primarily supported by the University of Chicago Materials Research Science and Engineering Center, NSF-MRSEC program under award NSF-DMR 2011854. The US Department of Energy, Office of Science, Basic Energy Sciences, under grant DE-SC0020972 provided support for TEV.





%

\end{document}